\documentclass[oldversion]{aa}
\usepackage{txfonts}
\usepackage{natbib}
\usepackage{graphicx}
\bibpunct{(}{)}{;}{a}{}{,}% magic!
\begin{document}
\headnote{Research Note}
\title{Magnetic flux 
	in the  inter-network quiet Sun from comparison
	with numerical simulations}
   \author{J.~S\'anchez~Almeida}% \and ...}
    \institute{Instituto de Astrof\'\i sica de Canarias, 
              E-38205 La Laguna, Tenerife, Spain}
   \offprints{J.~S\'anchez Almeida,
    \email{jos@iac.es}}
   \date{Received~~~~~/ Accepted~~~~~}
\abstract{
	\citeauthor{kho05} estimate the  mean magnetic
	field strength of the quiet Sun to be  20~G. The figure is 
	smaller than
	several existing estimates, and it comes from the
	comparison between observed Zeeman polarization signals
	and synthetic signals from numerical 
	simulations of magneto-convection.
	The numerical simulations require an artificially 
	large magnetic diffusivity, which smears out 
	magnetic structures smaller than the grid scale.
	Assuming a turbulent cascade for the unresolved
	artificially smeared magnetic fields, 
	we find that their unsigned
	magnetic flux is at least as important as 
	that explicitly shown in the simulation.  
	The unresolved fields do not produce Zeeman 
	polarization but contribute to the unsigned
	flux.
	Since they are not considered by 
	\citeauthor{kho05}, their mean magnetic
	field strength has to be regarded as a 
	lower limit.
	This kind of bias is not specific of a 
	particular numerical simulation or a spectral line. 
	It is to be expected when observed quiet Sun Zeeman 
	signals  are compared with synthetic signals 
	from simulations.
	}
\maketitle

\section{Rationale}\label{intro}

The observational properties of the quiet Sun magnetic
fields are not well established 
\citep[see, e.g., ][and references therein]{san04}.
We are in the way of characterizing them, 
a process whose end is difficult to predict partly
due to contradictions among existing measurements.
In this sense, \citet{kho05} estimate a mean field
strength or  
unsigned flux density $\langle B\rangle$ of only 
20~G, a value smaller than the unsigned flux
densities inferred by other means 
\citep[e.g.][]{fau01,dom03a,san03,tru04,bom05,
dom06}. 
This {\em Note} argues that the 
unsigned flux estimated by \citeauthor{kho05} should
be regarded as 
a lower limit rather than an unbiased
estimate.   If the conclusion is correct,
it may reconcile the
result of \citeauthor{kho05} with other
existing estimates claiming 
$\langle B\rangle $ to be of the order of 100~G.
Moreover, the argumentation and the caveat it entails
are of general nature.
They must be considered
whenever the quiet Sun unsigned flux is derived by
comparison of observed Zeeman signals with numerical
simulations.
Such approach to study the quiet Sun magnetism
is bound to play a major role in the future.

The paper by \citet{kho05} compares the polarization
observed in a quiet Sun region with the synthetic 
polarization produced by realistic numerical
simulations of solar magneto-convection. The 
numerical simulations have  a mean magnetic field
strength $\langle B\rangle$ that decreases exponentially
with time $t$, 
\begin{equation}
\langle B(t)\rangle\simeq
\langle B(0)\rangle\,
\exp(-t/\tau_D),
\label{efolding}
\end{equation}
with an e-folding time scale
$\tau_D\simeq 50\min $ 
(Khomenko \& V\"ogler 2005, private communication).
As usual, the angle brackets denote volume average.
The snapshot of the simulation 
with $\langle B(t)\rangle\simeq 20~$G is the one that best
reproduces the histograms of observed polarization
signals. This agreement is used to indicate that the 
observed inter-network region has $\langle B\rangle$
close to 20~G.
There is a natural
way to reconcile \citet{kho05} result with the
much larger fluxes  from other measurements. 
The  numerical 
simulations do not provide information on structures
smaller than a few grid points. They miss small scale 
magnetic structures  that
would show up if the simulations had finner
more realistic resolution.  The simulations only include
a part of the  magnetic fields, and so, it is to be expected 
that the estimate by \citet{kho05} only provides  
a lower limit to the true solar $\langle B\rangle$. 
This consideration can be a serious caveat or an
academic remark  depending
on the importance of the neglected 
magnetic fields. 
There is no definite way of evaluating 
the importance of the bias except for
improving the spatio-temporal resolution of the
simulations to reach realistic values.  
This approach, however, demands
a computer power exceeding by orders of magnitude
the current resources.
Our work provides a preliminary
estimate  
using the tools available at present.

\section{Energy and unsigned flux density
in the unresolved magnetic fields}

The simulations assume a magnetic diffusivity 
$\eta_t=1.1\times 10^{11}~$cm$^2$~s$^{-1}$, which
is some three orders of magnitude larger than
the Ohmic diffusivity to be expected at the
base of the photosphere 
\citep[$\eta\simeq 10^{8}~$cm$^2$~s$^{-1}$; ][]{kov83}.
This unrealistically high diffusivity is needed
for technical reasons to stabilize the numerical 
MHD code, and to prevent the artificial built-up of 
energy at the grid scale  
\citep{vog03b,vog05,nor90}.
	How can the numerical simulations be realistic 
	if they employ such artificially large magnetic 
	diffusivity? They can if there is 
	a physical mechanism that diffuses away 
	the magnetic energy at the rate imposed by the 
	artificial diffusivity. 
	Somehow the existence of such mechanism is implicit 
	when the numerical simulations are used to represent
	and characterize the real Sun. Lacking of obvious
	alternatives, we will describe such mechanism as  
	turbulent diffusion, 
%
%However, such diffusivity has a clear
%meaningful physical basis if it represents
%turbulent diffusion, 
similar to that 
required for the astrophysical dynamos to 
operate within reasonable time scales 
\citep[e.g., ][ Chapter 17]{par79}.
This association allows us to carry out
simple estimates. The turbulent diffusion
is produced by   
a complex unresolved magnetic
field with spatial scales 
$l$ so small that the diffusion
time scale for the true small Ohmic 
diffusivity  is similar to the diffusion 
time scale for the structures in the simulation 
\citep[][]{par79,bis03}.
Assuming that the high magnetic diffusivity
used by the numerical code is of turbulent
nature, one can evaluate the magnetic energy
in the turbulent magnetic field implicit in the 
simulation. This complex magnetic field 
does not produce polarization signals, 
so that the observables remain as those synthesized 
by \citet{kho05}. Nevertheless,
it contains energy and
unsigned flux that must be included
when estimating the true unsigned
flux density corresponding to the observed 
polarization signals. The section aims at showing how
such turbulent magnetic field can indeed 
contain a significant amount of energy and
unsigned flux.

In order to estimate the magnetic energy in the
turbulent cascade from the minimum resolvable
scale $L$, to the diffusive scale $l$, we need
a model for the MHD turbulence. 
The MHD turbulence is a topic of active research
and there is no unique and final way to approach
the problem 
\citep[see, e.g.,][and references therein]{bol05,bra05}.
However, one can estimate the magnetic energy in the
turbulent cascade using a few approximate
prescriptions which are now in use.
They all lead to the conclusion that this
energy may be significant.
 
We start off by assuming a Kolmogorov spectrum
for the turbulent cascade
\citep[e.g.,][]{bis03}. Then the total
magnetic energy per unit mass
corresponding to a (spatial) wavenumber
$k$ is,
\begin{equation}
E_k=C_K\,\varepsilon^{2/3}\,k^{-5/3},
\label{kolmo}
\end{equation}  
with
\begin{equation}
2\pi/ L< k < 2\pi/ l.
\end{equation}
The symbol $\varepsilon$ stands for the energy  
dissipated per unit time and unit mass by the turbulent cascade 
\citep[see, e.g.,][\S~5.3.2]{bis03}.
The Kolmogorov constant $C_K$ is a numerical
factor of the order of 1.7. Then the magnetic energy per unit mass
in unresolved fields
$E_t$ is,
\begin{equation}
E_t=\int_{2\pi/L}^{2\pi/l}E_k\,dk\simeq
	{{3C_K}\over{2(2\pi)^{2/3}}}(\varepsilon L)^{2/3},
	\label{eq4} 
\end{equation}
where we have taken into account that $l\ll L$. Now,
the mean magnetic field
of the numerical simulations have an exponential decrease
with an e-folding time scale $\tau_D$ (equation~[\ref{efolding}]).
Assuming that
the standard deviation among the field strengths
scales with the mean field\footnote{A condition
typical of many probability density functions suitable
for describing the  variable $B$, e.g.,
a Maxwellian distribution, and exponential distribution,
a uniform distribution, etc.},
\begin{equation}
\sigma_B=\langle(B-\langle B\rangle)^2\rangle^{1/2}
	\propto  \langle B\rangle,
	\label{sigma_stuff2}
\end{equation}
then,
\begin{equation}
\langle B^2\rangle=\sigma_B^2+\langle B\rangle^2
	\propto \langle B\rangle^2,
	\label{sigma_stuff}
\end{equation}
implying a magnetic energy 
$\langle B^2\rangle/(8\pi)$
decreasing exponentially
with a time scale half $\tau_D$
(compare equations~[\ref{efolding}] and [\ref{sigma_stuff}]).
The energy dissipated per unit time
by the turbulent cascade turns out to be 
\begin{equation}
\varepsilon\simeq 
	-{1\over{\rho}}\,{{d}\over{dt}}
	\big[{{\langle B^2\rangle}\over{8\pi}}\big]
	={{\langle B^2\rangle}\over{8\pi}} 
	2\,\tau_D^{-1}\,\rho^{-1},
\end{equation}
with $\rho$ representing the mass density
required to  transform  energy per unit
mass ($E_t$ and $E_k$) into energy per unit volume 
$\langle B^2\rangle/(8\pi)$. 
The dissipated energy has a compact expression
in terms of the Alfven speed $v_A$,
\begin{equation}
\varepsilon\simeq v^2_A\,\tau_D^{-1},
\label{eq6}
\end{equation}
with,
\begin{equation}
v_A^2={{\langle B^2\rangle}\over{4\pi\rho}}.
\end{equation}
If the symbol $B_t$ stands for the turbulent
magnetic field strength,
\begin{equation}
E_t={{\langle B_t^2\rangle}\over{8\pi\rho}}=
	{{\langle B_t^2\rangle}
	\over{\langle B^2\rangle}}
	{{v_A^2}\over{2}}.
\label{this}
\end{equation} 
Using equations~(\ref{eq4}), (\ref{eq6}), and (\ref{this}),
\begin{equation}
	{{\langle B_t^2\rangle}
	\over{\langle B^2\rangle}}=
	{{3C_K}\over{(2\pi)^{2/3}}}\Big({{L}\over{\tau_D\,v_A}}\Big)^{2/3}.
	\label{3ck}
\end{equation}
The variables defining our problem are,
$\rho\simeq 3\times 10^{-7}~$g~cm$^{-3}$ (the mass
density at the base of the photosphere),
$\langle B^2\rangle^{1/2}=28$~G (assuming
$\sigma_B\simeq \langle B\rangle\simeq 20~{\rm G}$), $L\simeq 100~$km
(Appendix~\ref{appa}), and
$\tau_D=50~$min (equation~[\ref{efolding}]). They render,
\begin{equation}
	{{\langle B_t^2\rangle}
	\over{\langle B^2\rangle}}\simeq 0.56.
\label{that}
\end{equation}
We are  interested in 
the mean magnetic field strength (or unsigned flux density),
since this parameter  is used
to characterize the
observations. Assuming the relationship~(\ref{sigma_stuff2})
with the same ratio $\sigma_B/\langle B\rangle$
for both  $\langle B\rangle$ and $\langle B_t\rangle$,
\begin{equation}
	{{\langle B_t\rangle}
	\over{\langle B\rangle}}
	=
	\sqrt{{\langle B_t^2\rangle}
	\over{\langle B^2\rangle}}\simeq 0.75.
	\label{this_other}
\end{equation}
The analysis above has been repeated using
the so-called Iroshnikov-Krishnan spectrum of MHD
turbulence \citep[][]{bis03},
\begin{equation}
E_k=C_{IK}\,(v_A\,\varepsilon)^{1/2}\,k^{-3/2}.
\label{iros}
\end{equation} 
(MHD turbulence
simulations seem to yield spectra
in between Kolmogorov and 
Iroshnikov-Krishnan; see \citealt{bol05}.)
Following the procedure explained above, one finds,
\begin{equation}
	{{\langle B_t^2\rangle}
	\over{\langle B^2\rangle}}=
	{{2 C_{IK}}\over{\sqrt{\pi/2}}}
	\Big({{L}\over{\tau_D\,v_A}}\Big)^{1/2}.
\label{ttt}
\end{equation} 
With $C_{IK}\simeq C_K$,  and for the numerical values used
in equation~(\ref{that}),
\begin{equation}
	{{\langle B_t^2\rangle}
	\over{\langle B^2\rangle}}\simeq 1.30,
\end{equation} 
yielding,
\begin{equation}
	{{\langle B_t\rangle}
	\over{\langle B\rangle}}\simeq 1.14.
\label{tttt}
\end{equation}
The estimates above assume the inertial range 
of the turbulence to begin at a scale larger 
than $L$. For this reason we can describe 
the magnetic energy for $k> 2\pi/L$ 
using laws characteristic of the inertial 
range  (equations~[\ref{kolmo}] and 
[\ref{iros}]). However, this assumption is not 
decisive. Consider a turbulent 
cascade whose inertial range begins at an
unresolved scale $\Lambda < L$.  Let us 
represent the unknown spectrum in the range between $L$ 
and $\Lambda$ as a power law of index $\alpha$,
\begin{equation}
E_k\simeq {{\langle B^2\rangle}\over{8\pi\rho}}
	\big({{L}\over{2\pi}}\bigr)^{1-\alpha}
	\,k^{-\alpha},
\end{equation}
with $2\pi/L < k \leq 2\pi/\Lambda$. The 
factor multiplying $k^{-\alpha}$ guarantees the 
continuity of  $E_k$ at $k=2\pi/L$; it is 
equal to the mean energy per unit mass and 
spatial frequency existing in the resolved 
scales, so that for $k=2\pi/L$,
\begin{equation}
E_k\simeq{{\langle B^2\rangle}\over{8\pi\rho}}
	{{1}\over{2\pi/L}}.
\end{equation}
The exponent $\alpha$ is also imposed by continuity 
arguments. When the inertial range begins, i.e. at  
$k=2\pi/\Lambda$, then $E_k$ is given by 
equation~(\ref{kolmo}). This constraint leads to
\begin{equation}
(\alpha-1)\,\ln(L/\Lambda) =
	\ln\big({{\langle B^2\rangle}\over{8\pi\rho}}\big)-
	\ln C_k-{{2}\over{3}}
	\ln\big({{\epsilon \Lambda}\over{2\pi}}\big).
	\label{crazy}
\end{equation}
The definition of $E_t$ in equation~(\ref{eq4}) implies,
\begin{displaymath}
E_t=
\int_{2\pi/L}^{2\pi/\Lambda} E_k\,dk+
\int_{2\pi/\Lambda}^{2\pi/l} E_k\,dk\geq
\int_{2\pi/L}^{2\pi/\Lambda} E_k\,dk\simeq
\end{displaymath}
\begin{equation}
~~~~~~~~~~~~~\simeq {{\langle B^2\rangle}\over{8\pi\rho}}\,
{{1}\over{\alpha-1}},
\label{myalpha}
\end{equation}
where we have assumed $\alpha> 1$ and $L \gg \Lambda$.
(Intermediate values of $\Lambda$ provide magnetic energies in 
between equation~[\ref{3ck}] and the limit considered here.) 
The expression~(\ref{myalpha}), together with equation~(\ref{this}),
render
\begin{equation}
\sqrt{{{\langle B_t^2\rangle}
	\over{\langle B^2\rangle}}}\ga {{1}\over\sqrt{\alpha-1}}
	\simeq 1.
	\label{add_fau}
\end{equation}
Once more the unsigned flux  in unresolved fields turns out 
to be significant.
The ratio in equation~(\ref{add_fau}) has been evaluated 
using equation~(\ref{crazy}) with the same parameters
leading to equation~(\ref{this_other}), which provide
an exponent $\alpha$ varying from 2.1 to 1.8 when 
$\Lambda$ goes from 10~km to 0.1~km. 

Equations (\ref{this_other}), (\ref{tttt})
and (\ref{add_fau})
suggest that the implicit 
turbulent field accounting for
the used magnetic diffusivity
has a magnetic flux similar to that explicitly 
shown in the simulations.

Some numerical simulations of turbulent magneto-convection 
show a range of wavenumbers 
where the magnetic energy exceeds the 
kinetic energy (super equipartition
range; see \citealt{bis03,mar04,bra05}). 
The magnetic energy densities used above are much smaller
than the kinetic energy density 
of the solar granulation. If this super 
equipartition range would have to be included 
in our estimate, one  needs to increase $\varepsilon$
with respect to the values employed above. 
The turbulent magnetic energy
would increase accordingly, leading to 
$\langle B_t\rangle$ larger than the values
in equations~(\ref{this_other})
and (\ref{tttt}). Consequently, 
the estimates above are probably conservative,
a conclusion reinforcing the importance of 
the unresolved magnetic fields. 

\section{Conclusions and discussion}
The  numerical simulations of solar 
magneto-convection require artificially
large magnetic 
diffusion, which smears out all the 
spectrum of magnetic structures smaller than
the grid scale. 
The use of  such  high 
magnetic diffusivity 
can be understood as the effect of a 
complex unresolved  turbulent magnetic
field with spatial scales 
so small that the diffusion
time scale for the true small Ohmic 
diffusivity is similar to the diffusion 
time scale for the structures in the simulation. 
Such an implicit magnetic field does not contribute 
to the polarimetric signals synthesized
by \citet{kho05}.
Assuming a turbulent cascade for the unresolved
artificially smeared magnetic fields, 
we find that their unsigned
magnetic flux is at least as important as 
that explicitly shown in the simulation.  
Should this magnetic flux is considered,
the Zeeman polarization signals measured by
\citeauthor{kho05}
are consistent with an unsigned flux of
$2\times 20~{\rm G}$
(i.e., $\langle B_t\rangle \simeq\langle B\rangle$;
equations~[\ref{this_other}], [\ref{tttt}]
and [\ref{add_fau}]
). 
In other words the unsigned flux assigned by
\citet{kho05}  must be regarded
as  a conservative lower limit.
This conclusion is not specific of the
simulations analyzed here. A  bias is to be
expected  whenever the quiet Sun unsigned
magnetic flux is inferred as the unsigned flux 
of numerical simulations reproducing 
observed Zeeman polarization signals.

The calculations that we describe represent
only a first approximation  to estimating the 
bias. They are based on the theory of MHD turbulence,
which remains to be completed.
The conclusions have to be  backed up or rejected 
by numerical simulations with realistic 
Ohmic diffusivities. 

%%%

\acknowledgements{
Thanks are due to E.~Khomenko and A.~V\"ogler for
details and clarifications on their work, and
to F.~Moreno-Insertis and
V.~Archontis for discussion on magnetic turbulence.
Thanks are also due to M. Faurobert for critical
and thoughtful comments on the original manuscript.
The work has partly been funded by the Spanish Ministry of Science
and Technology, 
project AYA2004-05792, as well as by
the EC contract HPRN-CT-2002-00313.
}

%%%%%%%%%%%%%%%%%

\appendix
\section{$L\simeq 100$~km}\label{appa}
The spatial discretization of the 
MHD equations solved by \citet{vog03b}
is based on a four-order finite difference 
scheme. Then the components of the magnetic
field are approximated
by a four-order polynomial, so that
the first and second  partial derivatives 
appearing in the MHD equations are represented
by a third-order polynomial and a second-order 
polynomial, respectively. We will estimate $L$
as the smallest 
wavelength of a sinusoidal magnetic field
for which the error involved 
in these polynomial approximations is 
insignificant.

Let us denote by $f(x)$ one of the components
of the magnetic field vector, with $k$ the wavenumber
along the spatial coordinate $x$,
\begin{equation}
f(x)=\sin(kx).
\end{equation}
The numerical code represents it 
in the surroundings of $x_0$ as
the polynomial $f_a(x)$,
\begin{equation}
f_a(x)=\sum_{n=0}^4f^{(n)}(x_0) {{(x-x_0)^n}\over{n!}}.
	\label{approximation}
\end{equation}
The fifth order term in the Taylor expansion of $f(x)$ 
provides  the error of the approximation, $\Delta f(x)$,
\begin{equation}
\Delta f(x)=f(x)-f_a(x)\simeq f^{(5)}(x_0) {{(x-x_0)^5}\over{5!}}.
\label{delta}
\end{equation}
As usual, the symbol $f^{(n)}$  denotes the
$n$-th derivative of $f(x)$.
The approximation~(\ref{approximation})
is carried out at each pixel, so that
\begin{equation}
|x-x_0|\leq \Delta x,
\label{rangeapp}
\end{equation}
with $\Delta x$ the pixel size. 
Among the various polynomials
used to represent the MHD equations, 
the computation of the second derivatives,
\begin{equation}
f^{(2)}(x)=-k^2\,f(x),
\end{equation}
corresponds to the lowest order  and, 
therefore, it is the worst approximation.
If the error associated 
with  the computation of the second derivatives is 
tolerable, then the polynomial approximation
is tolerable.
According to equation~(\ref{delta}), the error
$\Delta f^{(2)}(x)$ is
\begin{equation}
\Delta f^{(2)}(x)=f^{(2)}(x)-f_a^{(2)}(x)
	\simeq k^5\,\cos(kx_0)\, {{(x-x_0)^3}\over{3!}},
\end{equation}
where we have taken into account that
$f^{(5)}(x_0)=k^5\,\cos(kx_0)$.
The average (unsigned) error within the range
of the approximation (equation~[\ref{rangeapp}]) is,
\begin{displaymath}
\overline{|\Delta f^{(2)}(x)|}={{1}\over{2\Delta x}}
	\int_{x_0-\Delta x}^{x_0+\Delta x}
	|\Delta f^{(2)}(x)|\,dx
\end{displaymath}
\begin{equation}
	~~~~~~~~~~~~~~~~~\simeq
	k^5\,|\cos(kx_0)|{{\Delta x^3}\over{24}}. 
\end{equation}
The factor $|\cos(kx_0)|$ varies with $x_0$ but
its average is the same as the average of $|f(x)|$.
Then the mean relative error when computing the second
derivatives, $\delta$, is given by,
\begin{equation}
%{\overline{|\Delta f^{(2)}(x)|}
%\over{|f^{(2)}(x)|}}
\overline{|\Delta f^{(2)}(x)|}\big/
|f^{(2)}(x)|
=\delta
\simeq 
{{k^3\Delta x^3}\over{24}}.
\end{equation}
This  expression provides a relationship
between the relative error of the approximation,
the wavelength of the sinusoidal $\lambda$,
and the pixel size,
\begin{equation}
\lambda={{2\pi}\over{k}}\simeq  4.7\,\Delta x\,(\delta/0.1)^{-1/3}. 
\label{limitme}
\end{equation}
We define $L$ as the smallest wavelength for which the
error is insignificant, namely, $\lambda$
assuring $\delta\le 0.1$. 
According to equation~(\ref{limitme}),
and keeping in mind that $\Delta x= 21~$km, 
\begin{equation}
L\simeq 4.7\,\Delta x\simeq 100~{\rm km}.
\end{equation}

%
%\newcommand{\aap}{A\&A}
%\newcommand{\apj}{ApJ}
%\newcommand{\nat}{Nat.}
%\newcommand{\solphys}{Solar Phys.}
%\bibliography{/home/jos/texto/papers/sun}
%\bibliographystyle{apj}
%%
%\bibliography{/home/jos/texto/papers/sun}
%\bibliographystyle{apj}
%\input{ms.bbl}

\end{document}